\documentclass[12pt]{iopart}
\usepackage{graphicx}
\usepackage{amssymb}

\begin{document}

\title[Casimir torque]{Casimir torque between corrugated metallic plates}

\author{Robson B. Rodrigues$^{1}$, Paulo A. Maia Neto$^{1}$, Astrid Lambrecht$^{2}$ and Serge Reynaud$^{2}$}

\address{$^{1}$ Instituto de F\'{\i}sica, UFRJ, 
CP 68528,   Rio de Janeiro,  RJ, 21941-972, Brazil
\\
$^{2}$ Laboratoire Kastler Brossel,
CNRS, ENS, Universit\'e Pierre et Marie Curie case 74,
Campus Jussieu, F-75252 Paris Cedex 05, France
}

\begin{abstract}
We consider two parallel corrugated plates 
and show that a Casimir torque arises when the corrugation directions are not aligned. 
We follow the scattering approach and calculate the Casimir energy up to second order 
in the corrugation amplitudes, taking into account nonspecular 
reflections, polarization mixing and the finite conductivity of the metals. 
We compare our results with the proximity force approximation, which overestimates
the torque by a factor 2 when taking the conditions that optimize the effect. We argue that 
the Casimir torque could be measured for separation distances as large as 1 $\mu{\rm m}.$

\end{abstract}

\maketitle

\section{Introduction}

The 
relevance of the
Casimir effect \cite{Casimir} in connection with micro and nano-electromechanical systems (MEMS
and NEMS) has been recently highlighted \cite{Serry,Buks2001A,Buks2001B,Capasso_review,mexicanos}.     
 The attractive Casimir force
can lead to  permanent adhesion of the movable parts of MEMS and NEMS when
they are close enough, a phenomenon known as `stiction', resulting in
 malfunctioning of these devices. On the other hand, the Casimir effect 
may provide novel actuation schemes \cite{capassoA,capassoB} with promising potential applications.

Besides the usual normal Casimir force between metallic or dielectric
plates, the lateral Casimir force between corrugated plates \cite{chenA,chenB} can
also be used for micro-mechanical control. 
Very recently, two devices based on the lateral Casimir force were
theoretically proposed: a rack and pinion device \cite{golestanian},
 which is actuated by the lateral Casimir force between a 
cylinder with a corrugated surface and a corrugated plane plate, and a
Casimir ratchet \cite{emig},
driven by the lateral Casimir force between a plate with a symmetric
corrugation and a plate with an asymmetric corrugation. 

The experimental results for the lateral Casimir force 
were first compared with a theoretical analysis \cite{chenA,chenB}
based on the proximity-force approximation (PFA), or Derjaguin approximation \cite{DerjaguinA,DerjaguinB}.
 Within this approximation, the Casimir energy for non-planar surfaces is obtained by simply averaging the 
energy for parallel planes over the local separation distance. 
This approximation holds when the corrugation period $\lambda_C$ is much larger than the average 
separation distance $L,$ so that the surfaces are nearly plane in the scale of $L$ \cite{EPL2003}. 
It is extremely important to check the accuracy of this approximation, since it was  widely
employed for comparison with experimental results for the (normal) Casimir force 
between curved surfaces \cite{Ederth2000,ChenPRA04,DeccaAP05,Onofrio2005} (see 
Ref.~\cite{Reynaud_leipzig} for a more detailed discussion and a review on recent 
theoretical advances). 

We have computed the lateral force   beyond the PFA \cite{nonPFAlateralA,nonPFAlateralB} 
by employing the scattering approach \cite{NJP2006}, 
which takes into account the finite conductivity of the metallic plates as well as 
diffraction and polarization mixing. The corrugation is treated as a small perturbation of the plane geometry, 
and the lateral force is computed up to second order in the corrugation amplitudes $a_1$ and $a_2$ 
for each plate.
The perturbation expansion holds as long as $a_1,a_2\ll \lambda_C,L,$
but arbitrary relative values of $L$ and $\lambda_C$ are allowed, with the 
PFA regime corresponding to the limit $L/\lambda_C\rightarrow 0.$
This formalism
was also employed to compute the roughness correction to the Casimir force 
\cite{roughness}
and more recently
the lateral Casimir-Polder force \cite{Dalvit2007,Dalvit2007B}.
Beyond-PFA theories for uni-axial corrugation on
perfect reflecting plates were first reported by 
Emig {\it et al} for both perturbative \cite{Emig2003} and nonperturbative \cite{Emig2005} regimes.

The lateral Casimir force results from breaking the translational symmetry
along directions parallel to the plates \cite{Golestanian1997}. A more general situation occurs
when the corrugations are not aligned, so that 
the Casimir energy depends on the relative orientation between the two plates, and
 a Casimir torque arises \cite{EPL2006} (see figure~1). 
In this paper, we review the main physical properties of this effect.
Whereas the formalism developed in Refs.~\cite{Emig2003,Emig2005} requires the 
existence of a direction of translational symmetry (so as to allow for a convenient definition 
of field polarizations which are not coupled by the nonspecular reflection in this case), 
the scattering approach \cite{NJP2006,roughness} allows for a more general situation since it explicitly takes
 the coupling between different polarizations into account.
Thus, the scattering approach allows one to consider the geometry with rotated corrugated plates, as long 
as the corrugation amplitudes remain the smallest length scale as discussed above. 

Thanks to the high sensitivity of torsion balance techniques \cite{balance}, the Casimir torque 
between corrugated plates provides an attractive way to measure 
nontrivial ({\it i.e.} beyond-PFA) geometry effects. 
The use of torsion balances has also been proposed to measure the Casimir torque
between two (plane) birefringent  dielectric
plates \cite{Iannuzzi} (see also \cite{Parsegian,Barash,vanEnk,Torres}). 
For the proposed separation distances around
100 nm \cite{Iannuzzi}, the Casimir torque for corrugated plates 
is up to three orders of
magnitude larger than the torque between birefringent plates, for comparable values
of the separation distance and plate area, and taking realistic values for 
the corrugation amplitudes and the finite conductivity of the metallic plates. 
This could allow one to perform the experiment at larger 
separation distances, thus minimizing problems related to plate parallelism.

\section{Casimir energy for corrugated plates}

We assume that both plates have sinusoidal corrugation profiles with the same period $\lambda_C$
and amplitudes $a_1$ and $a_2.$
The corrugation lines of the bottom plate are along the $y$ direction ({\it ie,} the 
surface profile depends only on $x$). The top plate position along the $x$ axis is $b$
(by symmetry the energy does not depend on the position along the $y$ axis), and $\theta$ is 
the rotation angle. $b=0,\theta=0$ corresponds to the configuration where the corrugation lines are aligned 
with the surface crests facing each other (see figure~1). This is the configuration corresponding to 
the global energy minimum as discussed below.

\begin{figure}[t]
\includegraphics[width=8cm]{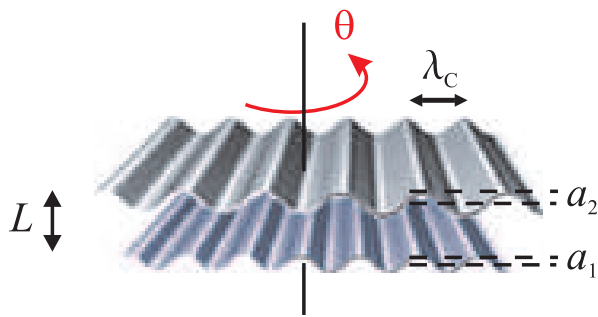}
\caption{Periodic corrugations
(period $\lambda_C$, amplitudes $a_1$ and $a_2$)
 are imprinted on both plates. $L$ is the average separation distance
and $\theta$ the rotation angle. We assume that $a_1,a_2\ll L, \lambda_C$.}
\end{figure}

\begin{figure}[t]
\includegraphics[width=9cm]{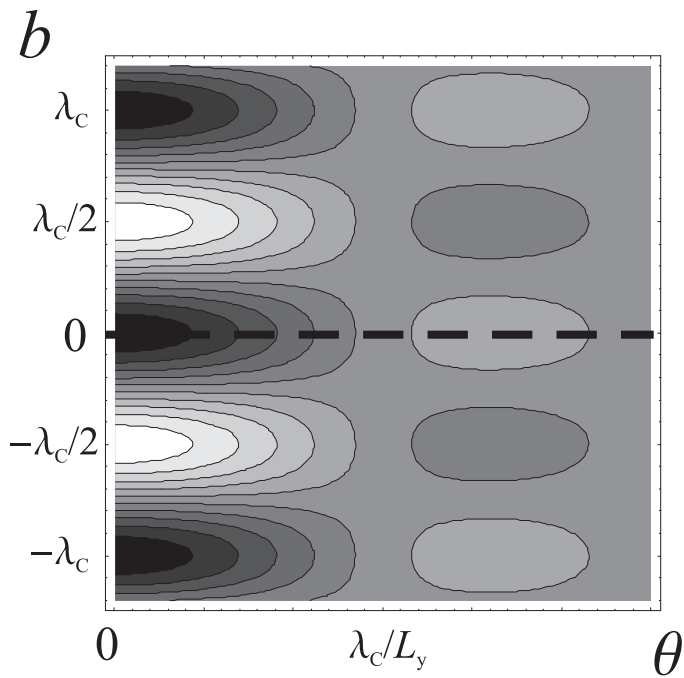}
\label{level}
\caption{
Level curves for the Casimir energy (in arbitrary units)
as a function of the lateral displacement $b$ and of the rotation angle $\theta.$
Regions with
lower energy values are darker.
}
\end{figure}

The top plate lateral dimensions are $L_x$ and $L_y,$ and $L$ is the average separation 
distance (along the $z$ direction). Both plates are assumed to be very large compared to
$L$ (so that border effects are negligible) as well as compared to $\lambda_C:$
$L_x \sim L_y\gg \lambda_C.$
The Casimir energy correction is then calculated to order $a_1a_2$ \cite{EPL2006}:   
\begin{equation}
\frac{\delta E_{\mathrm{PP}}}{L_{x}L_{y}}=\frac{a_{1}a_{2}}{2}\,\mathcal{G}%
(k)\cos \left( kb\right) \,\mathrm{sinc}(kL_{y}\theta /2).  \label{energy4}
\end{equation}%
where $k=2\pi/\lambda_C,$
$\textrm{sinc}(\xi)=\sin\xi/\xi$ 
 and $\mathcal{G}(k)$ is a response function which also depends on the separation distance $L.$
In figure \ref{level}, we plot the level curves of the Casimir energy
correction (in arbitrary units) as a function of $b$
and $\theta$. 
Since $\mathcal{G}(k)$ is always negative, the Casimir energy has  global minima at $\theta =0$ and $b=0$, $%
\lambda _{C}$, $2\lambda _{C}$, ... and local minima around $\theta \approx 1.43\lambda
_{C}/L_{y}$ (minimum of $\mathrm{sinc}(kL_{y}\theta /2)$) and $b=$ $\lambda
_{C}/2$, $3\lambda _{C}/2$, ...

If we start from $\theta =b=0$  and rotate the top plate around its center, we
follow the dashed line $b=0$ shown in figure \ref{level}. For $\theta
<\lambda _{C}/L_{y}$ the plate is attracted back to the minimum at $\theta =b=0$ without
sliding laterally. On the other hand, if the plate is released after a
rotation of $\theta >\lambda _{C}/L_{y}$ its subsequent motion will be a
combination of rotation and lateral displacement.
In the next section, we compute the Casimir restoring torque for the case of 
pure rotations with small rotation angles.

\section{Casimir torque}

The Casimir torque, given by 
\[
\tau =-\frac{\partial}{\partial\theta} \,\delta E_{\mathrm{PP}},
\]
is maximum at $\theta =0.66\lambda _{C}/L_{y}$ where it
is given by 
\begin{equation}
\frac{\tau }{L_{x}L_{y}}=0.109\,a_{1}a_{2}\,k\mathcal{G}(k)\,L_{y}.
\label{torque}
\end{equation}
As could have been expected, the torque per unit
area is proportional to the length of the corrugation lines $L_y$, which provides the scale for the 
moment arm.

We compute the response function ${\cal G}(k)$ using  
 the plasma model with plasma wavelength $%
\lambda _{P}=137$nm (corresponding to gold-covered plates).
In order to
fix the numbers given as examples below, we take
 $L_{y}=24\,\mu $m and corrugation amplitudes such that
$a_{1}a_{2}=200\,\mathrm{nm}^{2}$ (to be compared with $a_{1}a_{2}=472\,%
\mathrm{nm}^{2}$ in the lateral force experiment \cite{chenA,chenB}, where $a_{1}$
and $a_{2}$ were unequal).  Note that a change of these values
is easily taken into account by using the scaling law (\ref{torque}). 

If we also choose the corrugation period as in the lateral force experiment
($\lambda _{C}=1.2\,\mu \mathrm{m}$),
we find, 
at $L=100\,\mathrm{nm},$  $\tau
/(L_{x}L_{y})=5.2\times 10^{-7}\,\mathrm{N.m^{-1}}$, approximately three
orders of magnitude larger than the torque per unit area for birefringent
plates calculated in Ref.~\cite{Iannuzzi} for the most favorable
configuration at the same separation distance. The much larger figures found
in our case should allow one to perform the experiment at larger
separation distances.

At any given value of $L$, the torque between corrugated plates can be made
larger by choosing the corrugation period so as to maximize $k\mathcal{G}(k).$
For separation distances above 50 nm, this
corresponds to $k\approx 2.6/L$ or $\lambda _{C}\approx 2.4 L$. 
In figure 3, we plot the torque as a function of $k$ for $L=1\,\mu{\rm m}$
(solid line). The maximum at $k=2.6\,\mu{\rm m}^{-1}$ is indicated by
a vertical dotted line. We also show the values
obtained from the model with perfect reflectors (dashed line). They
overestimate the torque by $16\%$ near the peak region.

We recover the second-order PFA result for the Casimir torque from 
eq.~(\ref%
{torque}) by taking the limit $k\rightarrow 0$. The response function
satisfies the general `proximity force theorem' \cite{nonPFAlateralA} 
$\mathcal{G}(0)=e_{\mathrm{PP}%
}^{\prime \prime }(L),$ where $e_{\mathrm{PP}}$ is the Casimir energy per
unit area for parallel planes. We thus find: 
\begin{equation}
\left( \frac{\tau }{L_{x}L_{y}}\right) _{\mathrm{PFA}}=0.109\,a_{1}a_{2}%
\,ke_{\mathrm{PP}}^{\prime \prime }(L)\,L_{y}.  \label{torquePFA}
\end{equation}%
According to eq.~(%
\ref{torquePFA}), the torque grows linearly with $k$ in the PFA (dotted straight
line in figure 3). 
Figure 3 shows that the scattering curve  is very close to the 
PFA straight line when $k\ll 1/L$ as expected (${\cal G}(k)\approx {\cal G}(0)$). However, the 
discrepancy increases with $k,$ and at
 the peak value 
$k=2.6/L=2.6\,\mu \mathrm{m}%
^{-1}$ the PFA overestimates the torque by $103\%$.
 A detailed discussion of the ratio ${\cal G}(k)/{\cal G}(0)$ 
for several separation distances is presented 
in Ref.~\cite{nonPFAlateralB}.

\begin{figure}[t]
\includegraphics[width=12cm]{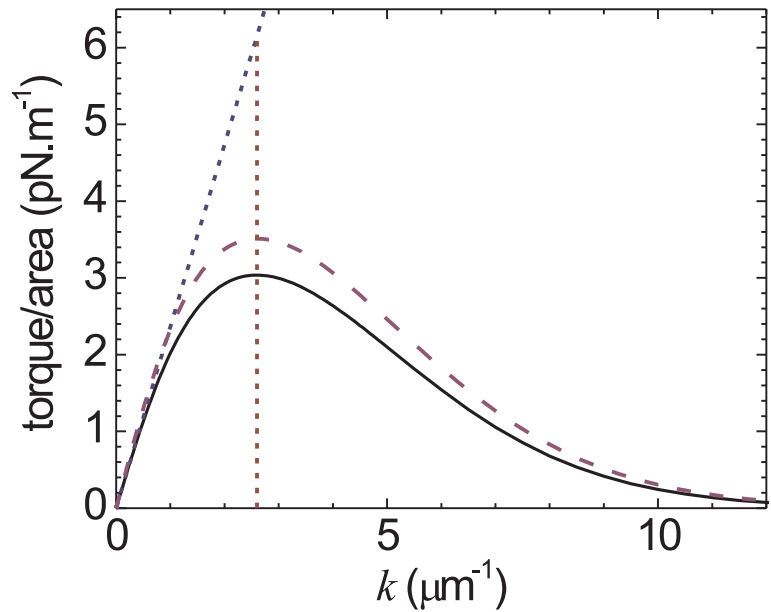}
\caption{
Torque as a function of $k=2\pi/\lambda_C$ for a separation distance $L=1\,\mu{\rm m}.$
Corrugation  amplitudes: $a_1a_2=200\,\mathrm{nm}^2.$ The plate length along the direction
of the corrugation lines is  
$L_y=24\,\mu\mathrm{m}.$ Results for gold-covered plates 
($\lambda_P=137\,\mathrm{nm}$) correspond to the solid (scattering) and 
dotted (PFA) lines; the dashed line corresponds to perfect reflectors. 
All results are computed up to second order in the corrugation amplitudes.
The vertical dotted line
indicates the optimal value $k=2.6\,\mu{\rm m}^{-1}.$ }
\end{figure}

\section{Conclusion}

As in the case of the lateral force, 
the Casimir torque 
between corrugated metallic plates
might have potential applications in the design of MEMS and NEMS.
We have studied the Casimir torque  with the help of the scattering approach, 
which provides an exact result for the second-order energy correction.

This torque is up to three orders of magnitude larger than the torque between
birefringent dielectric plates for comparable separation distance and area. 
The measurement of the Casimir torque with corrugated plates
would provide a direct demonstration of a non-trivial (beyond PFA) 
geometry dependence of the Casimir energy. 

RBR and PAMN thank FAPERJ, CAPES, CNPq
and Institutos do Mil\^enio de Informa\c c\~ao Qu\^antica e Nanoci\^encias
for financial support. AL acknowledges partial financial support 
by the European Contract STRP 12142 NANOCASE.

\end{document}